\def\comment#1{}
\newcommand{\Tr}{\mbox{Tr\,}}
\def\cc{{\rm c.c.}}
\newcommand{\tr}{\mbox{tr\,}}
\newcommand{\Sbar}{~\bar{}\!\!S}
\newcommand{\sla}[1]{{\hspace{1pt}/\!\!\!\hspace{-.5pt}#1\,\,\,}\!\!}
\def\lsim{\raisebox{-1.1mm}{$\,{\displaystyle\mathop{\sim}^{<}}\,$}}
\documentstyle[pra,epsf,aps]{revtex}

\def\cm#1{}

 \def\lfrac#1#2{{{{#1}/{#2}}}}

\newcommand{\sdag}{{\scriptsize \dagger}}

\begin{document}
\title{{
 Two Phase Transitions in Chiral Gross-Neveu Model
in  $2+ \epsilon$ Dimensions at Low $N$
}}
\author{H. Kleinert%
 \thanks{Email: kleinert@physik.fu-berlin.de~~URL:
http://www.physik.fu-berlin.de/\~{}kleinert \hfil
} and E. Babaev\thanks{Email: babaev@physik.fu-berlin.de
}}
\address{Institut f\"ur Theoretische Physik,\\
Freie Universit\"at Berlin, Arnimallee 14,
14195 Berlin, Germany}
\maketitle
\begin{abstract}
We show that the chiral Gross-Neveu model in $2+ \epsilon$ dimensions
has for a small number $N$ of fermions
two phase transitions corresponding to
{\em pair formation\/} and {\em pair condensation\/}.
 In the first transition,
fermions and antifermions
acquire spontaneously a mass
and are bound to pairs which behave
like a Bose liquid in a
chirally symmetric state.
In the second transition, the Bose liquid
condenses into a coherent state
which breaks chiral symmetry.
This suggests the possibility that
in particle physics,
the generation of quark masses
may also happen separately from the breakdown of chiral symmetry.
\end{abstract}

\section{Introduction}
The Nambu-Jona-Lasino (NJL) model  \cite{NJLM}
and its $N$-component version, the Gross-Neveu (GN) model \cite{GNM},
are field theories of zero-mass fermions
with quartic interaction
which provide
us
with considerable insight into the
mechanisms of spontaneous symmetry breakdown.
Both models can formally be turned
into pure boson theories.
In an SU($3$)$\times$ SU($3$)-symmetric version, the NJL model has has been shown to be equivalent to a
a
chirally SU($3$)$\times$SU($3$)
invariant $ \sigma$-model
which reproduces all well-known relations
of current algebra \cite{hadroniz}. For recent work and citations see \cite{handskog}.

The Gross-Neveu model
is exactly solvable
in the limit of $N\rightarrow \infty$.
For an attractive sign
of the interaction, a collective fermion-antifermion  field
acquires a nonzero vacuum expectation,
and the system shows quasi-long-range order.
In $D=2+ \epsilon$ dimensions,
the order of this state becomes proper long-range.
The ordered state is
reached
in a
second-order phase transition from a disordered state
if the
 renormalized coupling
constant $g$  becomes larger that a critical value $g^*=\pi \epsilon$.
The disordered state at small $g<g^*$ consists of massless
interacting fermions.
It exhibits chiral symmetry,
in which fermions are transformed by a phase rotation
containing a $ \gamma_5$ matrix.
In the ordered state at larger $g>g^*$, however,
the fermions acquire spontaneously a mass,
and the chiral symmetry
is broken spontaneously.

The purpose of this note is to point out, that
in a modified Gross-Neveu Model in which
pairs of fermions
form bound pair states analogous to the Cooper pairs in superconductor,
the paired phase becomes
incoherent in a Kosterlitz-Thouless-like transition
\cite{bkt}
if the number of field components
 $N$ drops below a certain critical value  $N_c\lsim 8$.
From this we conclude that
in $2+ \epsilon$ dimensions,
the transition
in which the pairs form
exists independently of a transition
in which they condense.

In the
ordinary Gross-Neveu model, the role of the
Kosterlitz-Thouless-like transition
is played by an Ising transition, which appears in addition
to the transition in which the collective state
forms.

\section{Proper Gross-Neveu Model}
The original Gross-Neveu model
has the following
O($N$)-symmetric Lagrange density
\begin{eqnarray} \label{8.67}
  {\cal L} = \bar\psi_a  i\sla{\partial}
       \psi _a + \frac{g_0}{2N} \left( \bar\psi _a
       \psi _a\right) ^2   .
\end{eqnarray}
where the index $a$ runs from $1$ to $N$.
At the mean-field level, the effective action is equal to the initial action
\begin{eqnarray} \label{8.86a}
  \Gamma \left[ \Psi ,\bar\Psi \right] = {\cal A} \left[
     \Psi ,\bar\Psi \right] = \int d^Dx \left[
       \bar \Psi  i \sla{\partial }
        \Psi + \frac{g_0}{2N} \left( \bar\Psi _a
         \Psi _a\right) ^2 \right]  .
\end{eqnarray}
In general, we obtain all Green functions
from the generating functional
\begin{eqnarray} \label{8.137}
{ Z[\eta , \bar\eta ] = e^{iW[\eta , \bar \eta ]}= }
  \int {\cal D} \psi {\cal D} \bar\psi
         e^{i{\cal A} \left[ \psi , \bar \psi \right]
            + i \left( \bar \psi \eta + \cc\right)
               },
\end{eqnarray}
where
 $\eta (x)$ and $\bar\eta (x)$
are fermionic anticommuting sources.
A collective field
$\sigma  \sim g \bar\psi \psi $ is introduced
to
rewrite (\ref{8.137}) as
\begin{eqnarray} \label{8.72}
    Z [\eta, \bar \eta ] & = & \int {\cal D} \psi {\cal D} \bar\psi
        {\cal D} \sigma e^{i \int d^Dx \left[ \bar\psi _a
                 \left( i \sla{\partial}  -\sigma \right)
                   \psi _a + (\bar\psi \eta + \cc)- {N } \sigma ^2/{2g_0}\right] }.
\end{eqnarray}
The fields $\psi(x) $ are integrated
out
according to the rule to
yield a generating functional containing only
the collective
field $ \sigma (x)$:
\begin{eqnarray} \label{8.73}
   Z [\eta, \bar \eta] = \int {\cal D} \sigma e^{i {\cal A}_{\rm coll}
       [\sigma ] - \bar \eta G_\sigma \eta},
\end{eqnarray}
with the collective action
\begin{eqnarray} \label{8.74}
  {\cal A}_{\rm coll} [\sigma ] = {N} \left\{ -
           \frac{N}{2g_0} \sigma ^2 - i \Tr \log
            \left[ i \sla{\partial}  - \sigma (x)\right]
            \right\}.
\end{eqnarray}
where Tr denotes the functional trace.

In the limit ${N} \rightarrow  {\infty} $,
the field $ \sigma $ is squeezed into the extremum  of the action
and we obtain the effective action
\begin{eqnarray} \label{8.75}
  \frac{1}{N} \Gamma \left[ \Sigma , \psi ,\bar\psi \right] & = &
  - \frac{1}{2g_0} \Sigma ^2 (x) - i \Tr \log \left[ i \sla{\partial}
      - \Sigma (x) \right]
 + \frac{1}{{N}} \Psi _a \left[ i \sla{\partial}-
 \Sigma (x) \right] \Psi _a
\end{eqnarray}
The extremum of $\Gamma \left[ \Sigma , \psi ,\bar\psi \right] $
 is given by the equtions of motion,
\begin{eqnarray} \label{8.76}
  \left[ i \sla { \partial}  - \Sigma (x) \right]
          \Psi _a (x) = 0,
\end{eqnarray}
\begin{eqnarray} \label{8.77}
  \Sigma (x) = g_0 \tr G_ \Sigma(x,x)
                -\frac{1}{N} g_0 \bar{\psi }_a \psi _a
                 (x)   ,~~~~G_ \Sigma(x,y)=
 \frac{i}{i \sla{\partial}
                 - \Sigma },
\end{eqnarray}
where the trace symbol $\tr $ is restricted to the Dirac indices.
The
expectation $\Psi_a $
of a fermionic field is always zero,
so that we only must solve the
  {\em gap equation\/}
\begin{eqnarray} \label{8.78}
  \Sigma (x) = g_0
\tr G_ \Sigma(x,x).
\end{eqnarray}
 Thus, as far as the extremum is concerned,
 we may study only the purely collective part of the exact
 action
\begin{eqnarray} \label{8.79}
  \frac{1}{N} \Gamma [\Sigma ]
           = - \frac{1}{2g_0} \Sigma ^2 - i \Tr \log
               i G_ \Sigma^{-1}.
\end{eqnarray}
The ground state is given by a constant gap field $ \Sigma_0$,
for which (\ref{8.78}) yields either $ \Sigma_0=0$
or
\begin{eqnarray} \label{8.80a1}
1 & = &  g_0     \,
 \tr  (1)\int \frac{d^D p_E}{(2\pi )^D}
                    \frac{1}{p_E^2 +  \Sigma _0^2}
 ,
\label{@gapeq}\end{eqnarray}
where we have performed
a Wick rotation
$p^0\rightarrow ip^4$ to
euclidean momenta
$p_\mu\equiv (p^1,p^2,p^3,p^4)$
with the metric $p_E^2=-p^2$.
The Dirac matrices have dropped out, except for the unit matrix
whose trace
is
$2^{D/2}$ for even $D$.
This expression may be continued to any non-integer value of $D$.

For a constant $\Sigma $, the effective action gives rise
to an effective potential
\begin{eqnarray} \label{8.81}
&&{  \frac{1}{{N}} v(\Sigma )
          = - \frac{1}{{N}} \Gamma [\Sigma ]
                    = }
 \frac{1}{2g_0} \Sigma ^2 - \tr (1) \frac{1}{2}
       \int \frac{d^D p_E}{(2\pi )^D} \log
       \left[ p^2_E +  \Sigma ^2\right].
\end{eqnarray}
Performing the integral yields in $D=2+ \epsilon$ dimensions
with $ \epsilon>0$
\begin{eqnarray} \label{8.84}
 \frac{1}{N} v(\Sigma )= \frac{\mu ^\epsilon }{2 }
        \left[ \frac{\Sigma ^2}{g_0\mu ^\epsilon }
           - b_\epsilon \left( \frac{\Sigma }{\mu }
             \right) ^{2 +  \epsilon  } \mu ^2\right],
\end{eqnarray}
where
$\mu $
is an arbitrary mass scale, and
the constant $b_\epsilon $ stands for
\begin{eqnarray} \label{8.85}
   b_\epsilon  =  \frac{2}{D} 2^{\epsilon /2} \Sbar_D
                   \Gamma (D/2) \Gamma (1 - D/2) =  \frac{2}{D} \frac{1}{(2\pi )^{D/2}}
               \Gamma (1- D/2) ,
\end{eqnarray}
which has an $ \epsilon$-expansion
   $b_\epsilon \sim -
            \left[ 1 - (\lfrac{\epsilon }{2}) \log \left( 2\pi  e^{-\gamma}
                \right) \right]/\pi \epsilon + {\cal O}(\epsilon ).$
A renormalized coupling constant $g$ may be introduced
by the equation
\begin{eqnarray} \label{8.87}
  \frac{1}{g_0 \mu ^\epsilon } - b_\epsilon \equiv \frac{1}{g},
\end{eqnarray}
so that
\begin{eqnarray} \label{8.89}
  \frac{1}{{N}} v(\Sigma ) = \frac{\mu ^\epsilon  }{2  }
           \left\{ \frac{\Sigma ^2}{g} + b_\epsilon \Sigma ^2
           \left[ 1 - \left( \frac{\Sigma }{\mu }\right) ^\epsilon
           \right] \right\}.
\end{eqnarray}
Extremizing this we obtain
either $ \Sigma_0=0$ or
a nonzero $ \Sigma_0$ solving
the gap equation
(\ref{8.80a1}) in the form
\begin{eqnarray} \label{8.90}
  1-\frac{g^*}{g} =  \frac{D}{2} \left(
         \frac{\Sigma _0}{\mu } \right) ^\epsilon,
\end{eqnarray}
where $g^*=-1/b_ \epsilon\approx \pi \epsilon$.
A nontrivial solution of this
is called  {\em gap\/}.
It specifies
the
mass which the fermions
acquire from
 the attractive interactions, and will
be denoted by $M$.
The second derivative of $v( \Sigma)$ shows that
the solutions
$ \Sigma_0=0$ and
$ \Sigma_0\neq0$ are stable  for
$g_0>0$ and
$g_0<0$, respectively.
Denoting the solution $ \Sigma_0$ of (\ref{8.90})
for $g=\infty$ by $M_\infty $, we may write the $g$-dependence of the gap as
\begin{equation}
 M(g)=M_\infty\left(1-\frac{g^*}{g}\right)^{1/ \epsilon}.
\label{@8.93p}\end{equation}
In terms of $M$, the effective potential (\ref{8.89}) can be rewritten as
\begin{eqnarray}
    \frac{1}{N} v(\Sigma ) =
-\frac{1}{4 \pi\epsilon}M^{D}
             \left[ D\!\left( \frac{\Sigma }{M}\right) ^2
             -2 \left( \frac{\Sigma }{M}\right) ^D\right]\!.
\label{8.132}\end{eqnarray}
It has a minimum at $ \Sigma=M$, where it yields
 the condensation energy
$    v(M ) =
-{N}M^D/4\pi$.
%

\section{Correlation Functions of Pair Field}
If $N$ is no longer infinite,
the pair field $ \sigma$
in the partition function
(\ref{8.72}) performs fluctuations around the extremal value $ \Sigma_0=M$.
For large $N$, the correlation functions
of $  \sigma(x)$ can be extracted from the
leading effective action
 (\ref{8.75})
 at $\Sigma _0=M$.
Setting $ \Sigma(x)= M+ \Sigma'(x)$,
we expand
\begin{eqnarray} \label{8.106}
  \delta ^2 \Gamma = - \frac{{N}}{2} \left[
          \frac{\Sigma'{}^2}{g_0} + i ~\Tr
           \left( \frac{i}{i\sla{\partial} - M}
            \Sigma ' \frac{i}{i \sla{\partial}-M}
            \Sigma '\right) \right],
\end{eqnarray}
implying a propagator of the $ \sigma'$-field
\begin{eqnarray} \label{8.107}
  G_{\sigma '\sigma '} = - \frac{1}{{N}}
         \frac{i}{\lfrac{1}{g_0} + \Pi (q)},
\end{eqnarray}
where $ \Pi (q)$ is given by the self-energy diagram
\begin{eqnarray} \label{8.108}
 \Pi (q)
      & = & - 2^{D/2} \int \frac{d^Dk_E}{(2\pi )^D}
                \frac{k(k-q)_E - M^2}
                {\left( k^2_E + M^2\right) \left[ (k-q)^2_E
                + M^2\right] },
\nonumber \end{eqnarray}
th integral being performed over euclidean $D$-dimensional energy-momenta.
With standard Feynman methods,
this can be transformed into a simple integral
\begin{eqnarray} \label{8.112}
  \Pi (q)
        & = & - \frac{D(D-1)}{2} b_\epsilon M^\epsilon \int^{-1}_{0}
              dx \left[ \frac{q_E^2}{M^2} x (1-x)+1\right] ^{
                 \epsilon /2}.
\end{eqnarray}
Inserting here
Eq.~(\ref{8.90}), we obtain
\begin{eqnarray} \label{8.113d}
 \!\!\!\!\!\!\!\!\!\!\!\!\!\!\!\!\!\!\!\frac{1}{g_0}+\Pi(q)
=
 \mu ^{\epsilon }
        \left(
\frac{1}{g^*}-\frac{1}{g}\right)\left\{-1+(D-1)
            \int^{1}_{0} dx
             \left[
\frac{q_E^2}{ M^2} x(1-x)
             +1\right] ^{\epsilon /2}
\right\},
\end{eqnarray}
which
can be expanded
for small $q$ as
\begin{eqnarray}
 \!\!\!\!\!\!\!\!\!\!\!\!\!\!\!\!\!\!\!\frac{1}{g_0}+\Pi(q)
=
         \epsilon \mu ^{\epsilon }\left(\frac{1}{g^*}-\frac{1}{g}\right)\left[1+(D-1)\frac{ 1}{12}
\frac{q_E^2}{ M^2} +\dots~
 \right] \label{@smallgq}\end{eqnarray}
This shows that the propagator (\ref{8.107})
has a correlation length
\begin{equation}
\xi=\left(\frac{D-1}{12 M^2}\right)^{1/2}.
\label{@corrole}\end{equation}
Inserting the $g$-dependence of $ M$ from (\ref{@8.93p}),
we see that
\begin{equation}
\xi=\frac{1}{M_\infty }\left(\frac{D-1}{12 }\right)^{1/2}\left(1-\frac{g^*}{g}\right)^{-1/ \epsilon}
\label{@crep}\end{equation}
so that
the coherence length
diverges for $g\rightarrow g^*$ with a
critical exponent
$ \nu=1/ \epsilon$.

\subsection{Chiral and Complex Pair Field Version of Model
with Goldstone Bosons}

We want to prove the existence of two
phase transitions in the chiral version of the Gross-Neveu model,
whose  Lagrange density is
\begin{eqnarray} \label{8.67b}
  {\cal L} = \bar\psi_a  i\sla{\partial}
       \psi _a + \frac{g_0}{2N}
\left[
\left( \bar\psi _a       \psi _a\right) ^2
+\left( \bar\psi _a  i \gamma_5     \psi _a\right) ^2
\right]  .
\end{eqnarray}
The
collective field action
(\ref{8.74}) is then replaced by
\begin{eqnarray} \label{8.74b}
  {\cal A}_{\rm coll} [\sigma ] = {N} \left\{ -
           \frac{N}{2g_0} (\sigma ^2+\pi^2) - i \Tr \log
            \left[ i \sla{\partial}  - \sigma (x)-i \gamma_5\pi\right]
            \right\}.
\end{eqnarray}
This model is invariant under the continuous set of chiral O(2)
transformations which rotate $ \sigma$ and $ \pi$ fields into each other.
This model is equivalent to yet
another one
which is closely related to the BCS model of superconductivity. Its Lagrange density is
\begin{eqnarray}
            {\cal L} = \bar{\psi}_a  i \sla\partial
       \psi _a + \frac{g_0}{2N} \left( \bar \psi _a C
        \bar\psi _a^T\right) \left( \psi _b^TC\psi _b\right).
\label{8.143}\end{eqnarray}
 Here $C$ is the
matrix of charge conjugation which is defined by
\begin{eqnarray}
                 C\gamma ^\mu C^{-1} = -\gamma ^{\mu T}.
\label{8.144}\end{eqnarray}
In two dimensions, we  choose the $ \gamma$-matrices
as
  $\gamma ^0  =
 \sigma^1,
~\gamma ^1  =
 -i\sigma ^2$,
and $C=\gamma ^1.$
Note that
$            \left( \bar\psi _a C \bar\psi_ a^T\right) ^\sdag
          = \psi^T_a C \psi _a$,
implying that $g_0 <0$ corresponds to an attractive potential.
The second model goes over into the first by replacing
$
  \psi \rightarrow  \frac{1}{2}(1-\gamma _5) \psi +  \frac{1}{2}(1+\gamma _5){2}
          C\bar\psi ^T,
$ where superscript T denotes transposition.
In the Lagrange density (\ref{8.143}) we introduce a complex
collective field by adding
a term $
   (\lfrac{{N}}{2g_0} )\left|  \Delta - \frac{g_0}{{N}}
 \psi ^T_b C \psi _b\right| ^2,$ leading to the partition function
\begin{eqnarray}
   \!\!\!\!\!\!\!\!\!\!\!\!
\!\!\!\!\! Z[\eta,\bar\eta] = \int {\cal D} \psi {\cal D} \bar\psi
   {\cal D} \Delta
        \exp\left\{ i \int d^Dx \left[ \bar\psi _a i\sla{\partial }
          \psi _a + \frac{1}{2} \left( \Delta ^\sdag \psi _a^T
          C \psi _a^\sdag + \cc\right)  + \bar\psi \eta + \bar\eta
         \psi  - \frac{N}{2g_0} | \Delta |^2\right]  \right\} .
\label{8.48}\end{eqnarray}
The relation with the previous collective fields
$ \sigma$ and $\pi$ is $ \Delta= \sigma+i \pi$.
In order to integrate out the Fermi fields we rewrite the free part of
Lagrange density in the matrix form
\begin{eqnarray}
     \frac{1}{2} \left( \psi ^T C,\bar\psi \right)
          \left( %
\begin{array}{cc}
     0  & i\sla\partial    \\{}
     i\sla\partial   & 0
\end{array}  \right)
     \left(
\begin{array}{c}
      \psi   \\{}
       C \bar\psi ^T
\end{array}\right)
\label{8.149}\end{eqnarray}
 which is the same as $\bar \psi  i \sla\partial
 \psi $, since
$ \psi ^T CC\bar\psi ^T =
 \bar\psi \psi
 ,~
    \psi ^T C\!\! \stackrel{\leftrightarrow }{\sla\partial }\!\! C \bar\psi ^T
 = \bar\psi\!\!  \stackrel{\leftrightarrow }{\sla\partial }\!\! \psi$.
But then the interaction with $\Delta $ can be combined with
(\ref{8.149}) in the form
%
  $\frac{1}{2} \phi ^T_i G_\Delta ^{-1} \phi$,
%
 where
\begin{eqnarray}
   \phi = \left( %
\begin{array}{c}
      \psi    \\{}
      C\bar\psi ^T
\end{array}\right)
,~~ \phi ^T = \left( \psi ^T, \bar\psi  C^{-1}\right)
\label{8.152}\end{eqnarray}
are doubled fermion fields, and
%
\begin{eqnarray}
    iG_\Delta ^{-1} = \left(
\begin{array}{cc}
    C  &  0  \\{}
    0   & C
\end{array} \right)
 \left( %
\begin{array}{cc}
    \Delta    &   i \sla\partial   \\{}
      i\sla\partial  &  \Delta ^\sdag
\end{array} \right)  = -\left( iG_\Delta ^{-1}\right) ^T
\label{8.153}\end{eqnarray}
is the inverse propagator in the presence of the external field $\Delta $.
Now we perform the functional integral over the fermion fields,
and obtain
\begin{eqnarray}
    Z[j] = \int {\cal D}\Delta {\cal D} \Delta ^\sdag
        e^{{iN}{\cal A} [\Delta ] + \frac{1}{2} j_a^T
           G_\Delta j_a},
\label{8.155}\end{eqnarray}
where ${\cal A}[\Delta ]$ is the collective action
\begin{eqnarray}
  {\cal A}[\Delta ] = - \frac{1}{2} |\Delta |^2 - \frac{i}{2}
      \Tr \log i G_\Delta ^{-1}
\label{8.156}\end{eqnarray}
and $j_a$
is the doubled version of the external source
\begin{eqnarray}
   j = \left(
\begin{array}{c}
    \bar\eta^T    \\{}
     C^{-1}\eta
\end{array} \right)      .
\label{8.157}\end{eqnarray}
This is chosen so that
$
  \bar\psi \eta + \bar\eta \psi  = \frac{1}{2}
           \left(j^T \phi - \phi ^T j\right)$.
In the limit $N \rightarrow {\infty} $, we obtain
from (\ref{8.155})
 the effective
 action
\begin{eqnarray}
      {\frac{1}{{N}} \Gamma [\Delta , \Psi ] =}
       \frac{1}{2g_0} |\Delta| ^2 - \frac{i}{2} \Tr \log
               i G_\Delta ^{-1} + \frac{1}{{N}} \bar\Psi_a
                 i G_\Delta ^{-1} \Psi _a
\label{8.158}\end{eqnarray}
in the same way as in the last chapter
for the simpler model with a real $ \sigma $-field.

The ground state has $\Psi  = 0$, so that the minimum
of the effective action
implies for $\Delta_ 0$  either $ \Delta_0=0$ or
the gap equation
\begin{eqnarray}
 1 = \frac{{g_0}}{2} \Tr G_{\Delta_0},
\label{8.159}\end{eqnarray}
where we may assume  $\Delta_0$ to be real.
With the Green function
\begin{eqnarray}
  G_{\Delta_0 } (x,y) = \int
        \frac{d^Dp}{(2\pi )^D}  e^{-ip(x-y)} \frac{i}{p^2- \Delta _0}
            \left(
\begin{array}{cc}
     \Delta _{0}  &   \sla p    \\{}
     \sla p &  -\Delta _0
\end{array} \right)
\left(
\begin{array}{cc}
     C^{-1}   &  0  \\{}
     0   &  C^{-1}
\end{array}
\right)  ,
\label{8.160}\end{eqnarray}
 the gap equation (\ref{8.159}) takes the
same form
as
(\ref{@gapeq}):
\begin{eqnarray}
   {1} =g_0 \,\tr (1) \int
           \frac{d^Dp}{(2\pi )^D} \frac{1}{p^2+M^2},
\label{8.161}\end{eqnarray}
where we have again set
$  M \equiv  \Delta _0$.
The renormalization of the coupling constant
and of the effective potential
yields the same equation for $v( \Delta_0)=v(M)$ as before,
so that the previous stability discussion
for $g<g^*$ and $g>g^*$
holds also here.

Let us now study the propagator of the complete
$\Delta $-field. For small deviations
$\Delta ' \equiv  \Delta -\Delta _0$
away from the ground state value
 we find from
(\ref{8.158}) the quadratic term
\begin{eqnarray*}
{ \frac{1}{N} \delta ^2 \Gamma }
 = - \frac{1}{2} \left\{ \frac{|\Delta |^2}{g_0} + \frac{i}{2}
          \Tr \left[
             \left(
\begin{array}{cc}
    \Delta '{}^\sdag  &    \\{}
      & \Delta '
\end{array} \right)
    G_M    \left(
\begin{array}{cc}
    \Delta '{}^\sdag &    \\{}
    &   \Delta '
\end{array}   \right) G_M\right] \right\}.
\end{eqnarray*}
The second term in curly brackets may be written more explicitly as
\begin{eqnarray*}
&&\!\!\!\!
   \frac{i}{2} \left[ M^2 (\Delta '{}^2 + \Delta '{}^{*2})
    2^{D/2} \int \frac{d^Dk}{(2\pi )^D} \frac{i}{k^2-M^2}
   \frac{i}{(k-q)^2 - M^2}\nonumber \right .
\left.+ 2|\Delta '|^2 \int \frac{d^Dk}{(2\pi )^D}
      \frac{i}{k^{2}-M^2}  \frac{i}{(k-q)^2-M^2} \tr [\sla{k}
    (\sla{k}-\sla{q})]\right],
\end{eqnarray*}
and becomes
\begin{eqnarray*}
 && \frac{1}{2} \left\{ M^2 \left( \Delta '{}^2 + \Delta '{}^{\sdag 2}
          \right) \tilde{\Pi }( q^2_E/M^2)
 + 2 |\Delta '|^2 \left[ \Pi ( q_E^2/M^2)  - M^2
          \tilde{\Pi } ( q^2_E/M^2)\right] \right\},
\end{eqnarray*}
where
 $ \Pi ( q_E^2/M^2) $ is the previous
   self-energy (\ref{8.108}),
and
  $\tilde \Pi \left( q_E^2/M^2\right) $
is the function
\begin{equation}
 \tilde{\Pi } \left( q^2_E/M^2\right)
    =  i 2^{D/2} \int \frac{d^Dk}{(2\pi )^D}
        \frac{i}{k^2-M^2} \frac{ i}{(k-q)^2 - M^2}
 = -\frac{D}{2} b_\epsilon  (1-D/2) \int^{1}_{0} d^Dx
          \left[ q_E^2 x (1-x) + M^2\right] ^{\lfrac{D}{2}-2}.
\label{8.171}\end{equation}
 As a result, the action for the quadratic deviations
from $ \Delta _0=M$ can be written as
\begin{eqnarray}
  \frac{1}{N}\delta ^2 \Gamma  = -\frac{1}{2} \left[
        \left( \frac{1}{g_0} + A\right) |\Delta '|^2 + \frac{1}{2}
        B \left( \Delta '{}^2 + \Delta '{}^{*2}\right) \right],
\label{8.172}\end{eqnarray}
 with the coefficients
\begin{eqnarray}
    A  =  - \frac{D}{2} b_\epsilon  M^ \epsilon\left[ (D-1) J_1^\epsilon
             \left( q^2_E/M^2\right) - (D/2-1) J^\epsilon _2
             (q_E^2/M^2)\right],
 ~~~~   B  =  \frac{D}{2} b_\epsilon M^\epsilon
 (D/2-1)
             J^\epsilon _2 \left( q_E^2/M^2\right),
\label{@}
\end{eqnarray}
and the integrals
\begin{eqnarray}
  J_1^\epsilon (z) = \int_{0}^{1} d^Dx \left[ zx(1-x)+1\right] ^{
                     \lfrac{D}{2} - 1}
 ,~~~
  J_2^\epsilon (z) = \int_{0}^{1} d^Dx \left[ zx(1-x)+1\right] ^{
                     \lfrac{D}{2} - 2}
 .
\label{8.173}\end{eqnarray}
Thus the propagators of real and imaginary parts of the field
$ \Delta'$ are
\begin{eqnarray}
G_{\Delta{_{\rm re} '}{\Delta{_{\rm re} '}}}  =  - \frac{i}{N}
       \frac{1}{\lfrac{1}{g_0} + A + B}  \label{8.174},~~~~
G_{\Delta{_{\rm im} '}{\Delta{_{\rm im} '}}}
 =
    - \frac{i}{N} \frac{1}{\lfrac{1}{g_0} + A - B } \label{8.175}.
\end{eqnarray}
The excitation spectrum
is given by the zeros
of  the denominator functions
\begin{eqnarray}
 &&\!\!\!\!\!\!\frac{1}{g_0} + A + B        =
    \frac{D}{2} b_\epsilon  M^\epsilon
          \left[ 1 - (D-1) J_1^\epsilon  \left( q_E^2/M^2
           \right) \right] ,\\
 &&\!\!\!\!\!\!\frac{1}{g_0} + A - B        =
    \frac{D}{2} b_\epsilon  M^\epsilon  \left\{
          \left[ 1 - (D-1) J_1^\epsilon  \left( q_E^2/M^2
           \right) \right] +(D-2) J_2^\epsilon
            \left( q_E^2/M^2\right) \right\}.
\label{@8.180'}\end{eqnarray}
By expanding $J_1^ \epsilon(z),~J_2^ \epsilon(z)$
 in powers of  $z = q_E^2/M^2 \approx 0$,
\begin{eqnarray}
  J_1^\epsilon(z)   \sim   1 +\frac{D-2}{12} z + {\cal O} (z^2),
  ~~~~J_2^\epsilon(z) \sim  1 + \frac{D-4}{12} z + {\cal O}(z^2)   ,
\label{8.182}\end{eqnarray}
  we find
\begin{eqnarray}
 \!\!\!\!\!\!\!\!\!\!\!\!\!\!\!\!\!\!\!
  \frac{1}{g_0} + A + B  =
 - \frac{D}{2} b_ \epsilon M^ \epsilon (D-2)
\left(1+\frac{D-1}{12}z\right)
                     + {\cal O}(z^2),
\label{8.183}
   ~~~\frac{1}{g_0} + A - B
              =  - \frac{D}{2} b_ \epsilon M^ \epsilon \frac{D-2}{12}
                     3z+ {\cal O}(z^2),
\label{8.183}\end{eqnarray}
Inserting here the gap equation (\ref{@8.93p}),
we obtain
\begin{eqnarray}
  \!\!\!\!\!\!\!\!\!\!\!\! \frac{1}{g_0} \!+\! A \!+\! B  =
 \epsilon\mu^ \epsilon\left(\frac{1}{g^*} -\frac{1}{g}\right)
\left(1+ \displaystyle\frac{D-1}{12}
                     \frac{q_E^2}{M^2}\right)+ \dots~,\label{8.183pa}
 ~~~~~  \frac{1}{g_0} \!+\! A \!-\! B  =
 \epsilon\mu^ \epsilon\left(\frac{1}{g^*} -\frac{1}{g}\right) \frac{1}{4}
                     \frac{q_E^2}{M^2}+ \dots~.
\label{8.183p}\end{eqnarray}
Recalling
(\ref{@smallgq}) we see that
the
propagator of $ \Delta'_{\rm re}$ coincides with that of $ \Sigma'$ in
the
standard Gross-Neveu model, so that
the fluctuations of $ \Delta_{\rm re} '$
have the same correlation length (\ref{@corrole}),
with at a critical exponent $ \nu$ in
(\ref{@crep}) as $g$ approaches $g^*$.

In contrast, the propagator of the
imaginary part of $ \Delta'$
has now a pole at $q^2=0$:
\begin{eqnarray}
 &&
G_{\Delta{_{\rm im} '}{\Delta{_{\rm im} '}}}
=   \frac{1}{N} \frac{4}{ \epsilon}
         \left(\frac{1}{g^*} -\frac{1}{g}\right)^{-1}
 M^2 \,\frac{i}{q^2} + \mbox{regular part at }
               q^2 =0.
\label{8.184a}\end{eqnarray}
The
 sign of the pole term
guarantees
 a positive
norm of the
 corresponding particle state in the Hilbert space.
The particle is a Nambu-Goldstone boson.

\section{Second Phase transition}

We are now prepared to show that the
pair version of the chiral Gross-Neveu model in $2+\epsilon$ dimensions
has two phase transitions.
Consider first the case $ \epsilon=0$
where
the collective field theory
consists
of complex field $ \Delta$
with O($2$)-symmetry $ \Delta\rightarrow e^{i\phi} \Delta$.
From the work of Kosterlitz and Thouless \cite{bkt}
we know
that such a field system
possesses
 macroscopic excitations
of the form of
 vortices and antivortices.
These attract each other by a logarithmic
Coulomb potential, just like a gas of electrons an positrons in two dimensions.
At low temperatures,
the vortices and antivortices
form bound pairs.
The grand-canonical ensemble of pairs exhibits
quasi-long-range correlations.
At some
 temperature $T_c$,
the vortex pairs break up, and the correlations
becomes short-range.
The phase transition
is of infinite order.

This transition is most easily understood in
a
model field theory involving of a pure phase field $\theta(x)$,
with a Lagrange density
\begin{equation}
{\cal L}=\frac{ \beta}{2}[\partial \theta(x)]^2,
\label{@modelld}\end{equation}
where  $ \beta$ is the stiffness of the $\theta$-fluctuations.
The important feature of the phase field $\theta$ is that it is
a cyclic field with $\theta=\theta+2\pi$.
In order to ensure that such jumps by $2\pi$
carry no energy,
the gradient in the Lagrange density needs a modification
which  allows for the existence of vortices and antivortices.
This will not be discussed here in detail,
 since the reader may consult the literature for it \cite{cmab,GFCM}.
We only state here that after including vortices and antivortices
at positions $x_i,~x_j$,
their partition function can be written as
\begin{equation}
Z=\sum _{\rm gas}\exp\left\{{4\pi^2 \beta}
\sum _{i<j}{q_i}{q_i}\frac{1}{2\pi}\log(|x_i-x_j|/r_0)\right\} ,
\label{@}\end{equation}
where $r_0$ is the size of the vortices.
For a single vortex-antivortex pair,
the average square distance $r^2$ is
\begin{equation}
<r^2>\propto \int_{r_0}^\infty dr r \,r^2 e^{-2\pi  \beta\log (r/r_0)}
\propto \frac{1}{2\pi  \beta-4}.
\label{@}\end{equation}
This diverges as the stiffness falls below $
 \beta_{\rm KT}={2/\pi}\approx 0.63662 .
$
A more detailed study shows that this
is an exact result for
a very dilute system of vortices and antivortices \cite{GFCM}.

The large-stiffness state
with bound vortex pairs
has a coherent phase field $\theta(x)$,
the low-stiffness state with separated vortex pairs
exhibits incoherent phase fluctuations.
The same situation is found in three dimensions,
only that the
excitations are
vortex lines.
These become infinitely long
and prolific in a second-order phase transition  \cite{GFCM}
at a critical point
$
 \beta_{c}\approx 0.33. $

The result (\ref{8.184a}) for $ \epsilon=0$
can now be used to estimate a critical value of the number of field components
$N=N_c$
below which the phase fluctuations of the complex field $ \Delta'$
become so violent that the system has a phase transition.
For this we write $ \Delta_{\rm im}'=M\theta$
and find from
(\ref{8.184a})
a propagator  of the $\theta$-field
\begin{eqnarray}
G_{\theta\theta}
 & \approx &
 \frac{i}{N}\frac{4\pi}{q^2}+{\rm regular~ terms}.
\label{8.190xx}\end{eqnarray}
Comparing this
with the
propagator
for the model Lagrange density  (\ref{@modelld})
\begin{equation}
G_{\theta\theta}
  =
\frac{1}{ \beta} \frac{i}{q^2}
\label{@propcpomp}\end{equation}
we identify the stiffness $ \beta=N/4\pi$.
The
pair version of the chiral Gross-Neveu model
has therefore a vortex-antivortex pair breaking transition
if $N$ falls below the critical value
$
N_c=8.$

Consider now the model
in $2+ \epsilon $
dimensions
where pairs form
at $g=g^*\approx \pi \epsilon$.
A comparison between
the propagator
(\ref{8.184a}) and (\ref{@propcpomp})
yields a stiffness of phase fluctuations
\begin{equation}
 \beta
=\frac{N}{4\pi}  \left(1 -\frac{g^*}{g}\right).
\label{@stiffn}\end{equation}
The linear vanishing of the stiffness with the distance
of the coupling constant $g$ from the critical value
${g^*}$
is in agreement with a general scaling relation,
according to which the critical exponent of
bending rigidity should be equal to $(D-2) \nu$.
Since the model has $ \nu=1/ \epsilon$ [see (\ref{@crep})],
this yields $(D-2) \nu=1$,
which is precisely the exponent in Eq.~(\ref{@stiffn}).

\begin{figure}[tb]
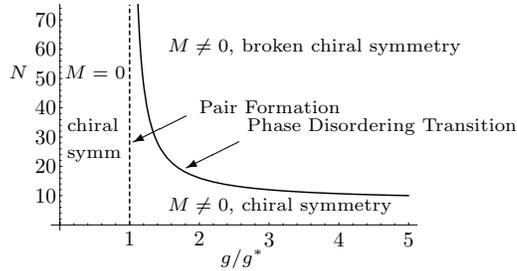

~~~~~~~~~~~~~~~
~~~~~~~~~~~~~~~
\input ncofg.tps
\caption[]{The two transition lines in the $N-g$-plane
of the chiral Gross-Neveu model in $2+ \epsilon$ dimensions.
For $ \epsilon=0$, the vertical transition line coincides with the $N$-axis,
and the solid hyperbola degenerates into a horizontal line at $N_c=8$.
The quark masses and chiral properties are indicated.
}
\label{ncofg.tps}\end{figure}
The stiffness (\ref{@stiffn})
implies the existence of a phase transition
in the neighborhood of two
and in three dimensions
at
roughly
\begin{equation}
N_c\approx 8
 \left(1 -\frac{g^*}{g}\right)^{-1},~~~D\approx2,~~~~
~~~~~~~N_c\approx 4.19
 \left(1 -\frac{g^*}{g}\right)^{-1},~~~D=3.
\label{@}\end{equation}
As $N$ is lowered below these critical values,
the phase fluctuations of the pair field $ \Delta $ become
incoherent and the pair
condensate dissolves.
The different phases are indicated in
  Fig.~\ref{ncofg.tps}.
In the chral formulation of the same model, the intermediate phase has
chiral
symmetry
in spite of a nonzero spontaneously generated ``quark mass" $M\neq 0$.
The reason why this is possible is that
the ``quark mass"
depends only on $| \Delta_0|$,
thus allowing for arbitrary phase fluctuations
preserving chiral symmetry.

The sceptical reader may wonder
whether the solid hyperbola in
  Fig.~\ref{ncofg.tps}
is not simply the proper (albeit approximate)
continuation of the
vertical line for smaller $N$.
There are two simple counterarguments.
One is formal: For infinitesimal $ \epsilon$
the first transition lies precisely at $g=g^*=\pi \epsilon$
for {\em all\/} $N$,
so that the horizontal transition line is clearly distinguished
from it. The other argument is physical.
If $N$ is lowered at some very large $g$, the binding energy of the
pairs {\em increases with $1/N$\/}
\{in
two dimensions,
the binding energy
is
$4M\sin^2[\pi/2(N-1)]\}$.
It is then
impossible that the
phase fluctuations on the horizontal branch of the transition line,
which are low-energy excitations,
unbind the strongly bound pairs.
This will only happen in the limit $N\rightarrow \infty$
where the binding energy becomes zero and the two transition
curves merge into a single curve.
This is the situation
in
the BCS theory of superconductivity, where Cooper pair binding and pair condensation
coincide.

In the ordinary Gross-Neveu model, the
analog of the phase disordering
transition is an Ising transition, in which the vacuum expectation value of $ \sigma$
jumps between $-  \Sigma_0$ and $ \Sigma_0$ in a disorderly fashion.
In two dimensions, this occurs at some
critical value  $N_c$.
In $2+ \epsilon$ dimensions,
this transition should again exist independently of the transition at which
the system enters into a state of nonzero $ \Sigma_0$.
It will  be interesting to
see these two transitions in either model
confirmed by Monte-Carlo simulations.

\end{document}